\documentclass[12pt]{article}

\pdfoutput=1

\usepackage{cite}

\usepackage{amssymb,latexsym}

\usepackage{epstopdf}

\usepackage{graphics,epsfig}

\usepackage{color}
\usepackage{xcolor}

\usepackage{amsmath,amsfonts}

\usepackage{mathrsfs}

\DeclareMathAlphabet{\mathpzc}{OT1}{pzc}{m}{it}

\usepackage{feynmp}

\let\a=\alpha \let\b=\beta \let\g=\gamma \let\d=\delta \let\e=\epsilon
\let\z=\zeta  \let\th=\theta  \let\k=\kappa
\let\l=\lambda \let\m=\mu \let\n=\nu \let\x=\xi \let\p=\pi 
\let\s=\sigma   \let\f=\phi  \let\y=\psi
 
        \let\Th=\Theta \let\L=\Lambda
\let\X=\Xi  \let\S=\Sigma  \let\Y=\Psi
 
\let\la=\label  
  
\def\nn{\nonumber} \def\bd{\begin{document}} \def\ed{\end{document}}
\def\ds{\documentstyle} \let\fr=\frac \let\bl=\bigl \let\br=\bigr
\let\Br=\Bigr \let\Bl=\Bigl
\let\bm=\bibitem
\let\na=\nabla
\def\tU{{\widetilde U}}
\let\pa=\partial \let\ov=\overline
\def\ie{{\it i.e.\ }}
\newcommand{\be}{\begin{equation}}
\newcommand{\ee}{\end{equation}}
\def\ba{\begin{array}}
\def\ea{\end{array}}
\def\ft#1#2{{\textstyle{{\scriptstyle #1}\over {\scriptstyle #2}}}}
\def\fft#1#2{{#1 \over #2}}
\def\F#1#2{{ F_{#1}^{(#2)} }}
\def\cF#1#2{{ {\cal F}_{#1}^{(#2)} }}

\def\R{{\bf R}}
\def\sst#1{{\scriptscriptstyle #1}}
\def\oneone{\rlap 1\mkern4mu{\rm l}}
\def\e7{E_{7(+7)}}
\def\td{\tilde}
\def\wtd{\widetilde}
\def\im{{\rm i}}
\def\bog{Bogomol'nyi\ }
\newcommand{\ho}[1]{$\, ^{#1}$}
\newcommand{\hoch}[1]{$\, ^{#1}$}
\newcommand{\bea}{\begin{eqnarray}}
\newcommand{\eea}{\end{eqnarray}}
\newcommand{\ra}{\rightarrow}
\newcommand{\lra}{\longrightarrow}
\newcommand{\Lra}{\Leftrightarrow}
\newcommand{\ap}{\alpha^\prime}
\newcommand{\bp}{\tilde \beta^\prime}
\newcommand{\cB}{{\cal B}}
\newcommand{\cO}{{\cal O}}
\newcommand{\vecx}{\vec{x}}
\newcommand{\vecy}{\vec{y}}
\newcommand{\vecp}{\vec{p}}
\newcommand{\vecq}{\vec{q}}
\newcommand{\tr}{{\rm tr} }
\newcommand{\Tr}{{\rm Tr} }
\newcommand{\NP}{Nucl. Phys. }

\newcommand{\cL}{{\cal L}}
\newcommand{\cA}{{\cal A}}
\newcommand{\cT}{{\cal T}}
\newcommand{\cR}{{\cal R}}
\newcommand{\cD}{{\cal D}}
\newcommand{\cH}{{\cal H}}

\def\Cb{\bar{C}}

\def\sst#1{{\scriptscriptstyle #1}}
\def\0{{\sst{(0)}}}
\def\1{{\sst{(1)}}}
\def\2{{\sst{(2)}}}
\def\3{{\sst{(3)}}}
\def\4{{\sst{(4)}}}
\def\5{{\sst{(5)}}}
\def\6{{\sst{(6)}}}
\def\7{{\sst{(7)}}}
\def\8{{\sst{(8)}}}
\def\9{{\sst{(9)}}}
\def\p{{\sst{(p)}}}
\def\q{{\sst{(q)}}}
\def\ve{\varepsilon}
\def\vf{\varphi}
\def\F{\Phi}
\def\wg{\wedge}

\def\thb{\bar{\theta}}
\def\Thb{\bar{\Theta}}
\def\barp{\bar{p}}
\def\barq{\bar{q}}
\def\barc{\bar{c}}
\def\bard{\bar{d}}
\def\e{\epsilon}

\def \bi{\bibitem}
\def \la {\label}

\def \l {\lambda}
\def\foot{\footnote}
\def \tl  {{\tilde \l}}
\def \sql {{\sqrt \l}}
\def \adss {$AdS_5 \times S^5$\ }
\newcommand{\rf}[1]{(\ref{#1})}
\def \ov {\over}

\def\th{\theta}
\def\Th{\Theta}
\def\vth{\vartheta}
\def\btheta{{\bar\theta}}
\def\ttheta{{{\tilde\theta}}}
\def\bttheta{{{\bar\ttheta}}}
\def\vth{\vartheta}

\def\ra{\rightarrow}
\def\N{\nabla}
\def\F{{\cal F}}
\def\uM{\underline{M}}
\def\uA{\underline{A}}
\def\uN{\underline{N}}
\def\uP{\underline{P}}
\def\ua{\underline{a}}
\def\ub{\underline{b}}
\def\uc{\underline{c}}
\def\ud{\underline{d}}
\def\ue{\underline{e}}
\def\uf{\underline{f}}
\def\ui{\underline{i}}
\def\uj{\underline{j}}
\def\uk{\underline{k}}
\def\ul{\underline{l}}
\def\ual{\underline{\alpha}}
\def\ube{\underline{\beta}}
\def\um{\underline{m}}
\def\un{\underline{n}}
\def\up{\underline{p}}
\def\uq{\underline{q}}
\def\ur{\underline{r}}
\def\us{\underline{s}}
\def\umu{\underline{\mu}}
\def\unu{\underline{\nu}}
\def\ula{\underline{\l}}
\def\uka{\underline{\k}}
\def\usi{\underline{\s}}
\def\urh{\underline{\r}}
\def\cc{\circ}
\def\eqv{\equiv}

\def\ni{\noindent}

\def\Ep{E^{{}^{(+)}}}
\def\Em{E^{{}^{(-)}}}

\def\Mp{M^{{}^{(+)}}}
\def\Mm{M^{{}^{(-)}}}

\def \ha{{1\ov 2}}

\def\r{\rho}

\def\Y{{\rm Y}}
\def\X{{\rm X}}
\def\tY{\tilde{\rm Y}}
\def\tX{\tilde{\rm X}}
\def\dY{\dot{\rm Y}}
\def\dX{\dot{\rm X}}

\def \J {\mathcal{J}}
\def \del {\partial}

\def\dF{\dot{F}}
\def\dG{\dot{G}}
\def\df{\dot{f}}
\def \E {{\cal E}}
\def \S {{\cal S}}
\def \J {{\cal J}}

\def\ms{\mathcal{S}}
\def\mj{\mathcal{J}}
\def\soj{\fr{\ms}{\mj}}
\def \R {{\bf R}}
\def \om {\omega}
\def \bE {\bar E}
\def \x {{\cal X}}

\def \bi{\bibitem}
\def \la {\label}

\def \l {\lambda}
\def\foot{\footnote}
\def \tl  {{\tilde \l}}
\def \sql {{\sqrt \l}}
\def \adss {$AdS_5 \times S^5$\ }
\def \ov {\over}

\def \varpi {{\rm w}}

\def\thb{\bar{\theta}}
\def\Thb{\bar{\Theta}}
\def\mb{\bar{\m}}
\def\ab{\bar{\a}}
\def\zb{\bar{z}}
\def\psib{\bar{\psi}}
\def\barp{\bar{p}}
\def\barq{\bar{q}}
\def\barc{\bar{c}}
\def\bard{\bar{d}}
\def\e{\epsilon}
\def\wb{\bar{w}}
\def\lb{\bar{\l}}
\def\Jb{\bar{J}}
\def\Nb{\bar{N}}
\def\Zb{\bar{Z}}
\def\pab{\bar{\pa}}

\def\At{\tilde{A}}
\def\Bt{\tilde{B}}
\def\Ct{\tilde{C}}
\def\Dt{\tilde{D}}
\def\Et{\tilde{E}}
\def\Ft{\tilde{F}}
\def\Gt{\tilde{G}}
\def\Ht{\tilde{H}}
\def\Kt{\tilde{K}}
\def\Mt{\tilde{M}}
\def\Nt{\tilde{N}}
\def\Rt{\tilde{R}}
\def\at{\tilde{a}}
\def\bt{\tilde{b}}
\def\ct{\tilde{c}}
\def\dt{\tilde{d}}
\def\et{\tilde{e}}
\def\ft{\tilde{f}}
\def \ztt{\tilde{\z}}
\def \zetat{\tilde{\zeta}}
\def\htil{\tilde{h}}
\def\gt{\tilde{g}}
\def\nt{\tilde{n}}
\def\mut{\tilde{\mu}}
\def\nut{\tilde{\nu}}
\def\pht{\tilde{\f}}
\def\Phit{\tilde{\Phi}}
\def\vft{\tilde{\vf}}

\def\rht{\tilde{\rho}}

\def\asth{\hat{*}}
\def\phh{\hat{\phi}}

\def\bA{{\bf A}}

\def\ola{\overleftarrow}
\def\ora{\overrightarrow}
\def\alt{\tilde{\a}}

\def\eh{\hat{e}}
\def\eph{\hat{\e}}
\def\ph{\hat{p}}
\def\alh{\hat{\a}}
\def\beh{\hat{\b}}
\def\gah{\hat{\g}}
\def\Fh{\hat{F}}
\def\muh{\hat{\m}}
\def\nuh{\hat{\n}}
\def\thh{\hat{\th}}
\def\rhh{\hat{\r}}
\def\dh{\hat{d}}
\def\ih{\hat{i}}
\def\jh{\hat{j}}
\def\hh{\hat{h}}
\def\nh{\hat{n}}
\def\gh{\hat{g}}
\def\kh{\hat{k}}
\def\deh{\hat{\d}}
\def\wh{\hat{w}}
\def\lah{\hat{\l}}
\def\Ah{\hat{A}}
\def\Gh{\hat{G}}
\def\Kh{\hat{K}}
\def\Nh{\hat{N}}
\def\Rh{\hat{R}}
\def\Ch{\hat{C}}
\def\Omh{\hat{\Omega}}

\def\xh{\hat{x}}

\def\ps{\rlap{\, /}\;\,p }
\def\ks{\rlap{\, /}\;\,k }

\def\gym{g_{YM}}

\def\adot{\dot{a}}
\def\bdot{\dot{b}}
\def\bpa{\bar{\pa}}

\def\pr{\prime}
\def\ssk{\medskip}
\def\clb{\color{blue}}
\def\clr{\color{red}}
\def\clg{\color{green}}
\def\clp{\color{purple}}
\def\clc{\color{cyan}}
\def\clm{\color{magenta}}
\def\cly{\color{yellow}}

\def\bfA{{\bf A}}
\def\bfB{{\bf B}}
\def\bfK{{\bf K}}
\def\bfU{{\bf U}}
\def\bfX{{\bf X}}
\def\bfY{{\bf Y}}
\def\bfZ{{\bf Z}}
\def\bfg{{\bf g}}
\def\bfn{{\bf n}}

\def\bsk{\bigskip}
\def\ssk{\medskip}

\def\Ec{{\cal E}}

\begin{document}

\overfullrule=0pt
\parskip=2pt
\parindent=12pt
\headheight=0in \headsep=0in \topmargin=0in
\oddsidemargin=0in

\vspace{ -3cm}
\thispagestyle{empty}

 \vspace{0.1cm}

\setcounter{equation}{0}
\setcounter{footnote}{0}
\setcounter{section}{0}

\begin{center}

{\Large\bf  Quantum-gravitational trans-Planckian energy of a time-dependent black hole}

\vskip 0.8cm

\vspace{0.5cm}

A. J. Nurmagambetov$\,^{\spadesuit}$\let\thefootnote\relax\footnotetext{$^{\spadesuit}$ Also at {\it Karazin Kharkov National University, 4 Svobody Sq., Kharkov, UA 61022} \& {\it Usikov Institute for Radiophysics and Electronics, 12 Proskura St., Kharkov, UA 61085}. }
and I. Y. Park{$^\dagger$}
\\

\vspace{0.3cm}

$^{\spadesuit}$
{\it Akhiezer Institute for Theoretical Physics of
NSC KIPT,\\
1 Akademicheskaya St., Kharkov, \\ UA 61108 Ukraine \\
ajn@kipt.kharkov.ua
}

\vspace{0.3cm}
{\it {}{$^\dagger$}Department of Applied Mathematics,
Philander Smith College 
                               \\
Little Rock, AR 72223, USA \\
inyongpark05@gmail.com
}

 \vspace{.5cm}

\end{center}

 \vspace{0.1cm}

\begin{abstract}

We continue our recent endeavor in which a time-dependent black hole solution of a one-loop quantum-corrected Einstein-scalar system was obtained and its near-horizon behavior was analyzed. The energy analysis led to a trans-Planckian scaling behavior near the event horizon.  In the present work the analysis is extended to a rotating black hole solution of an Einstein-Maxwell-scalar system with a Higgs potential. Although the analysis becomes much more complex compared to that of the previous, we observe the same basic features, including the quantum-gravitational trans-Planckian energy near the horizon.

\end{abstract}
\vspace{1in}
Key words: quantum gravity, loop effects, Firewall, trans-Planckian energy

\newpage





\section{Introduction}

We are entering an era of muti-messenger astrophysics, and a substantial amount of new data is being collected for various astronomical objects. It is by now firmly established that diverse ultra-high-energy cosmic rays (UHECRs) of extra-galactic origins constantly bombard Earth's atmosphere. Since the energy scale of these particles - $\sim 10^{19}$ eV - far exceeds that of LHC, study of their origin may well allow us to take a leap in solving some of as-yet unsolved problems in the field.

Although the accumulated data indicate that active galactic nuclei (AGNs) should be largely responsible for the generation of UHECRs, the precise mechanism is yet to be understood. There is wide consensus that the UHECRs must be the work of the super-massive black holes at the centers of the active galaxies. Therefore the focus of one's quest should be the physics that can produce various extreme-high-energy particles - such as gamma ray photons, protons, heavier ions, and neutrinos - in massive volume.  
Motivated by this and more theoretically-oriented issues, such as black hole information (see, e.g., \cite{Mathur:2009hf}\cite{Skenderis:2008qn} for reviews) and Firewall \cite{Almheiri:2012rt}\cite{Braunstein:2009my}, we have initiated in \cite{Park:2014mba,Park:2017dib,Nurmagambetov:2018het,AJNIYP:2018} the study of quantum gravitational effects as the potential agent behind certain astrophysical phenomena, including the generation of the UHECR particles.

It is conventionally believed that the quantum gravitational effects are largely negligible. (The same has been believed in the astrophysical situations such as in astrophysical black hole environs.) This view has been challenged through a series of recent works, according to which the quantum-gravitational effects may not only be observable but may also be behind some of the spectacular astrophysical phenomena. 
In this work we continue our recent endeavor in which a time-dependent black hole solution of a one-loop quantum-corrected Einstein-scalar system was obtained and its near-horizon behavior was analyzed. The energy analysis led to a trans-Planckian energy behavior near the event horizon, which should lead to observable effects. We extend the analysis to an Einstein-Maxwell-scalar system in the present work.

To put things in an orderly perspective, let us summarize the status of the matters surrounding the new gravity quantization approach proposed in \cite{Park:2014tia}. A gravity theory can be shown to be one-loop offshell renormalizable in the the conventional covariant quantization framework: for instance it was shown in the classic paper by 't Hooft and Veltman \cite{'tHooft:1974bx} that pure Einstein gravity is one-loop renormalizable. Once matter fields are included, the one-loop offshell renormalizability is lost. (However, the renormalizability is restored once one includes a cosmological constant that provides more leverage to absorb the one-loop ultraviolet divergences.) At two-loop, things become worse and the conventional offshell renormalizability is lost. Therefore, although one could conduct various quantum-level studies, the nonrenormalizability would force one to stipulate that the results be taken up to the issue of renormalizability, which has been a frustrating setback to further progress. 
Motivated by the holographical reduction of the physical states \cite{Park:2014tia},\footnote{A related discussion can be found in \cite{Hadad:2019lxw}.} the covariant quantization has recently been revisited \cite{Park:2016zgt}: a way out of the longstanding nonrenormalizability has been proposed, establishing the renormalizability of the {\em physical} states that are a certain subset of the perturbative offshell states. Explicit one-loop renormalization procedures have been worked out for several gravity-matter systems \cite{Park:2019amz}. 
With the new scheme of quantization, one can be assured that the one-loop analysis will remain valid even to higher loops. Further out, the development has provided a stage for further investigation of quantum-gravitational effects and their applications. Through our recent works it has been shown that the quantum-gravitational effects are of ``order-1" in the sense to be reviewed below.

What sets the present work apart from the previous works is that the system being considered is more realistic: we consider an Einstein-Maxwell-scalar system with a Higgs-type potential and its rotating black hole solution. As in \cite{Park:2017dib}\cite{Nurmagambetov:2018het}\footnote{See \cite{Kawai:2017txu}  for a related result:  there it was observed in a time-dependent setup that the quantum stress-energy tensor inside the black hole reaches a near-Planckian value.}, the analysis leads to a trans-Planckian energy. The trans-Planckian energy behavior may well be a  generic feature of a time-dependent black hole configuration at the quantum level.

\vspace{.3in}
The rest of the paper is organized as follows.
\vspace{.1in}

In section 2, before embarking on the technical analysis, we give a qualitative reasoning on why there ought to be a trans-Planckian energy behavior near the event horizon.  
In section 3, after reviewing the Einstein-scalar system analyzed in \cite{Nurmagambetov:2018het}, we obtain a time-dependent quantum-level solution of an Einstein-Maxwell-scalar system. We consider the $\L_0=0$ case - where $\L_0=0$ denotes the classical part of the cosmological constant - for a reason to be explained; extension to the $\L_0 \neq 0$ case is left for future. The classical part of the quantum-level solution is required to settle down to a Kerr geometry. A novel feature observed in \cite{Park:2016fxc,Park:2016vam} is shared: the quantum effects remove the time-dependence of the classical part, which is crucial for the subsequent energy analysis. 
In section 4, we analyze the energy observed near horizon by an infalling observer and are led to a trans-Planckian energy. 
Although the quantum-induced trans-Planckian energy may sound radical, the result is obtained within the norm of quantum field-theoretic techniques.\footnote{ 
Although the framework has radically new ingredients,  it is only two- and higher- loop renormalizability that requires such ingredients: the {\em one-loop} renormalizability can be established within the conventional framework. The subsequent techniques of finding the time-dependent solution and analyzing the curved space scalar electrodynamics are standard.} Toward the end of section 4, we briefly comment on the boundary conditions. We reason that the perfect infall boundary condition that is used in the context of the quasi-normal modes is rather restrictive and that more general boundary conditions must be considered to describe the physics of the ring-down phase of a black hole.
In section 5, which is the concluding section, we summarize the results and list future directions.

\section{Physical origin of trans-Planckian energy}

The present work is motivated in part by the Firewall proposal; let us briefly review the Firewall argument. The backbone of the Firewall argument is as follows. For simplicity let us take a Schwarzschild black hole. Consider the Kruskal observer and the corresponding vacuum. The Kruskal vacuum must not be a vacuum to a Schwarzschild observer and should appear to be radiating - which is nothing but the Hawking radiation - to a Schwarzschild observer. Now let us consider things in `reverse': consider a Schwarzschild observer and the corresponding vacuum {(or an eigenstate of the observer's number operator)}. Similarly as before, the Schwarzschild vacuum {(or the eigenstate)} must not be a vacuum to a Kruskal observer. What makes this part of the physics more dramatic is that the Kruskal observer is infalling so the radiation the observer will encounter is highly blue-shifted near the horizon, a Firewall.

One of the goals of the present work (and its sequels) is to back up the Firewall proposal by a quantitative analysis in a more realistic astrophysical environment. What we set out to check in the present work is conceptually simple but highly complicated technically: for one thing, we intend to calculate the quantum-corrected energy measured by an infalling observer near the horizon of a time-dependent rotating black hole of a Einstein-Maxwell-scalar system. To this end, one needs the quantum-corrected action and the field equation with its solution: in particular, a time-dependent solution. Afterwards, one needs to work out the four-velocity of the observer, $U^\m$, in the quantum-corrected background.  
The stress-energy tensor $T_{\m\n}$ is obtained by taking the functional derivative of the matter part of the action with respect to the metric. We quote the classical part of it here for convenience: 
\bea
{ T_{\m\n}} &=& - \fr2{\k^2}\L g_{\m\n}+g_{\m\n}\Big[-|\pa_\r \psi-iqA_\r \psi|^2 -m^2|\psi|^2 -\fr14 F_{\r\s}^2   \Big] \nn\\
&&\hspace{-.3in} +  \left[(\pa_\mu \psi-iqA_\m \psi)(\pa_\n \psi^*+iqA_\n \psi^*)+(\m \leftrightarrow \n)\right]+  F_{\m\r}F_\n{}^\r
+{\cal O}(\hbar),
\eea
where $g_{\m\n}, A_\r, \psi$ denote the metric, vector field, and scalar, respectively. 
The energy density $\r$ measured by the observer is given by $\r\equiv U^\m U^\n T_{\m\n}$ where the full quantum-level solution is to be substituted into $T_{\m\n}$. Finally, one evaluates the contributions from each term in $T_{\m\n}$. As we will see, some of the terms in $\r$ exhibit a trans-Planckian behavior, which is also an overall behavior of the entire energy density $\r$.

There is a clear qualitative way to see how the presence of the quantum correction part leads to a trans-Planckian energy (whereas the purely classical analysis doesn't): when computing the energy density, evaluate the right-hand side of the metric field equation, ${T}_{\m\n}=\fr{G{\m\n}}{8\pi G}$, and contract with the four-velocities. As we will see, the classical metric comes to take the same form as the classical Kerr form that yields a finite energy (zero energy, more precisely). Being an additional contribution to the stress-energy tensor, the presence of the quantum modes changes this status (as we will see in more detail) by directly analyzing the stress-energy tensor in section 4.

One may wonder about the physical origin of such a non-smooth structure in the vicinity of the horizon. Consider a particle heading toward the black hole. It will produce other particles through the quantum-field-theoretic chain reactions on its way to the event horizon.\footnote{In the conventional picture, pair creation process has been argued to be responsible for the Hawking radiation.  
	Although the pair-creation process is expected to be one of the main channels of the quantum-gravitational effects, our picture posits a much more complex process than the conventional one where one of the pair particles falls into the black hole while the other one escapes, thereby causing the information loss at the end.
	
	When the BH undergoes an active accretion, such effects must produce a sufficient amount of the outgoing flux for detection. In other words, the quantum-gravitational effects should lead to mass production of cascading particles, some of which escape the black hole.} As an infalling particle accelerates toward the black hole, the acceleration would increase without bound and the particle will release the energy though various radiation channels such as synchrotron radiation and bremsstrahlung.

One thing worth noting is that the quantum-gravitational effects introduce non-minimal coupling terms such as $\psi\psi^* R$. Although the effect of such non-minimal terms is small (and will not play a role in this work), they will lead to violation of the Equivalence Principle. To recap, on one hand, the quantum gravitational effects are responsible for the trans-Planckian energy, and on the other hand, they produce such non-minimal couplings, which are at odds with the classical-level understanding of the Equivalence Principle and/or should limit the range of validity thereof.

\section{Time-dependent solutions}

Although it is conventionally believed that the quantum gravitational effects are largely negligible, in general there are circumstances, as our recent works have revealed, in which the quantum-gravitational effects are of ``order-1". Such effects may be behind some of the highly energetic astrophysical phenomena. 
In the previous works \cite{Park:2017dib}\cite{Nurmagambetov:2018het}, we considered an Einstein-scalar system and examined the possibility that the quantum corrections may produce a violent energy behavior in the vicinity of the event horizon. 
It was shown that for a time-dependent solution, one indeed encounters such a behavior. In the present work we consider an Einstein-Maxwell-scalar system in extension of the previous works.

In section 3.1 we start by reviewing our earlier work of \cite{Nurmagambetov:2018het} on an Einstein-scalar case. We highlight some salient results of the analysis and point out an undesirable feature of the solution therein obtained. This motivates introduction of a potential in the scalar sector; we consider a Higgs-type potential. 
In section 3.2, a time-dependent solution of the Einstein-Maxwell-scalar system with a Higgs potential is obtained. 
For a reason we will detail, we set the classical part of the cosmological constant to zero, $\L_0=0$; extension to $\L_0 \neq 0$ case will be pursued elsewhere.

\subsection{Review of Einstein-scalar case}

Let us warm up by reviewing the case considered in \cite{Nurmagambetov:2018het}, in which an Einstein-scalar system was considered. 
The classical action of \cite{Nurmagambetov:2018het} is 
\bea
 S=\fr1{\k^2}\int d^4 x \sqrt{-g}\Big[R-2\L\Big] 
-\int d^4 x \sqrt{-g}\Big[\fr12(\pa_\mu \z)^2 +\fr12m^2\z^2\Big]. \la{csa}
\eea
It admits an AdS black hole solution, 
\bea
ds^2=-\fr1{z^2}\Big(Fdt^2+2dtdz \Big)+\Phi^2(dx^2+dy^2)   \la{cs}
\eea
with $\z=0$, and
\bea
F=-\fr{\L}{3}-2M z^3\quad,\quad \Phi=\fr1{z}. 
\eea
The one-loop 1PI effective action is given by \cite{Park:2016zgt} 
\bea
&&\hspace{.1in} S=\fr1{\k^2}\int d^4 x \sqrt{-g}\Big[R-2\L\Big] 
-\int d^4 x \sqrt{-g}\Big[\fr12(\pa_\mu \z)^2 +\fr12m^2\z^2\Big]\nn\\
&&\hspace{-.3in}+\fr1{\k^2}\int d^4 x \sqrt{-g}\Big[e_1{ \k^4} R\z^2+e_2 \k^2R^2+e_3\k^2 R_{\m\n}R^{\m\n}+e_4 { \k^6}(\pa\z)^4+e_5{ \k^6} \z^4+\cdots\Big],  \la{qsactscalar} \nn\\
\eea
where the $e$'s are numerical constants that can be determined with the chosen renormalization conditions. The quantum-level field equations can be obtained by varying the action of eq.\!\! \rf{qsactscalar}.\footnote{As a matter of fact, the boundary conditions must be considered before varying the action. We refer to \cite{Park:2019amz} for potential issues associated with the boundary conditions in varying a gravitational action.
} The quantum system admits the following form of the time-dependent metric solution \cite{Murata:2010dx}:
\bea
ds^2=-\fr1{z^2}\Big(F(t,z)dt^2+2dtdz \Big)+\Phi^2(t,z)(dx^2+dy^2)  \la{ma}
\eea
with the quantum-corrected series
\bea
\hspace{-.5in}F(t,z)&=& F_0(t) +F_1(t) z+ F_2(t)z^2+F_3(t)z^3 + ...\nonumber\\
&+&\k^2 \Big[F_0^h(t) +F_1^h(t) z+ F_2^h(t)z^2+F_3^h(t)z^3 + ...\Big], \nn\\
\Phi(t,z)&=&\frac{1}{z}+\Phi_0(t) +\Phi_1(t) z+ \Phi_2(t)z^2+\Phi_3(t)z^3 + ...\nonumber\\
&+& \k^2\Big[\fr{\Phi_{-1}^h(t)}{z}+\Phi_0^h(t) +\Phi_1^h(t) z+ \Phi_2^h(t)z^2+\Phi_3^h(t)z^3 + ...\Big]. 
   \la{1stans}  \nn\\
\eea
Similarly, for the scalar:
\bea
\z(t,z)&=&\z_0(t) +\z_1(t) z+ \z_2(t)z^2+\z_3(t)z^3 + ...\nonumber\\
&+&\k^2\Big[ \z_0^h(t) +\z_1^h(t) z+ \z_2^h(t)z^2+\z_3^h(t)z^3 + ...\Big], \la{zetaser}
\eea
where the modes with superscript `$h$' represent the quantum modes. The cosmological constant is set to 
\bea
\L\equiv \L_0+\hbar \k^2\L_1,
\eea
where $\hbar$ has been explicitly displayed for convenience. 
The order-by-order analysis of the field equations in $z$ and $\hbar$, for the classical modes, leads to
\bea
&&m^2=\fr{2\L_0}{3},\quad \z_0=0 ,\quad { F_0=-\fr{\L_0}{3}},\quad \Phi_1=0,\quad {F_1=-F_0 \Phi_0-\Lambda_0 \Phi_0}\,, \nn\\
&&\hspace{.5in}W_2=0,\quad F_2=\frac{1}{4} \left(4 F_0 \Phi_0{}^2-8 \pa_t \Phi _0 \right),\nn\\
&& \z_3=0,\quad \Phi_3=0  ,\quad {F}_3=\mathrm{const},\quad   \z_4=0,\quad \Phi_4=0 ,\quad  F_4=-F_3\Phi_0 \; ; 
 \la{mrone2}  \nn\\
\eea
and for the quantum modes, to 
\bea
&&\hspace{-.3in}\z_0^h=0,\quad { { F_0^h=-\frac{1}{3}  \Lambda_1}} ,\quad
\Phi_{1}^h=0,\quad { F_1^h}=\frac{2}{3} \Big(3 {F}_0^h \Phi_0+{\Lambda_0} \Phi_0 {\Phi}_{-1}^h-{\Lambda_0} {\Phi}_0^h-3 \pa_t{\Phi}_{-1}^h \Big), \nn\\
&&\Phi_2^h=0, \quad { F_2^h}=  \frac{1}{3} \Big(-{  {\Lambda_1} \Phi_0{}^2}+2 {\Lambda_0} \Phi_0{}^2 {\Phi}_{-1}^h -2 {\Lambda_0} \Phi_0 {\Phi}_0^h+6 {\Phi}_{-1}^h \pa_t{\Phi} _0-6 \pa_t{\Phi}_0^h  \Big),
\nn\\
&&\hspace{-.5in}\quad  { \z_3^h}=   -\fr1{\text{$\Lambda_0$}}  \Big({\text{$\Lambda_0 $} \text{$\zeta $}_1^h \Phi_0{}^2+ 2 \text{$\Lambda_0 $} \text{$\zeta $}_2^h \Phi_0+ 3 \Phi_0 \text{$\pa_t{\zeta} $}_1^h+ 3 \text{$\zeta $}_1^h \pa_t{\Phi} _0+  3 \text{$\pa_t{\zeta} $}_2^h} \Big)    ,     \quad   \Phi_3^h=0,\quad  { \pa_t{F}_3^h}=-3F_3\pa_t{\Phi}_{-1}^{h}\,,
\nn\\
&&\hspace{-.3in} F_4^h=F_3 \Phi_0 \Phi_{-1}^h-F_3 \Phi_0^h-F_3^h \Phi_0,\quad
\Phi_4^h=-3 {e_2} F_3 \Phi_0{}^2+3 {e_2} F_5-2 {e_3} F_3 \Phi_0{}^2+2 {e_3} F_5 \,,\nn\\
&&\hspace{-.5in} \z_4^h=  \frac{F_3 \zeta_1^h}{2 \text{$\Lambda_0 $}} +\frac{12 \pa_t{\zeta}_1^h \pa_t{\Phi} _0}{\Lambda_0^2}+\frac{6 \Phi_0 \pa^2_t{\zeta}_1^h}{\Lambda_0^2}+\frac{6 \zeta_1^h \pa^2_t{\Phi} _0}{\Lambda_0^2}+\frac{6 \pa^2_t{\zeta}_2^h}{\Lambda_0^2}+\frac{9 \Phi_0{}^2 \pa_t{\zeta}_1^h}{\text{$\Lambda_0 $}}+\frac{9 \Phi_0 \pa_t{\zeta}_2^h}{\text{$\Lambda_0 $}}+\frac{9 \zeta_1^h \Phi_0 \pa_t{\Phi} _0}{\text{$\Lambda_0 $}} \nn\\ 
&&\hspace{3in}+2 \zeta_1^h \Phi_0^3+3 \zeta_2^h \Phi_0^2.
 \la{mrtwo2}
\eea
The subsequent analysis then leads to a trans-Planckian energy for which the quantum modes played a crucial role (see \cite{Nurmagambetov:2018het} for more details; the corresponding analysis will be carried out in section 4). As noted in \cite{Park:2016fxc}, consideration of the field equations at the quantum level leads to additional constraints among some of the {\em classical} modes as well, and in particular yields
\be
\z_1=0=\z_2\,.
\ee
Because these two modes serve as the building blocks of the higher modes, the entire tower of the classical modes comes crumbling down.
This shows that the quantum-level field equations deform the classical part as well, although one may naively expect that the classical part will remain intact. Many features of the analysis in \cite{Nurmagambetov:2018het}, including the one just mentioned, are present in the Einstein-Maxwell-scalar system, as we will see.

The solution above has an undesirable feature. Because the quantum effects force the entire classical part of the scalar field to vanish, the classical part of the cosmological constant $\L_0$ must vanish as well. With $\fr1{\L_0}$ appearing in some of the mode relationships above, this can potentially be a problem. It turns out that this is not a genuine problem: one can set $\L_0=0$ from the beginning.\footnote{If one actually sets $\L_0=0$ from the beginning, one gets a different solution. This means that the classical limit approaches the usual Kerr as opposed to the dS/AdS Kerr.}
Not unrelated to this, one of the mode relations, $m^2=\fr{2\L_0}{3}$, does not look natural. These observations motivate introduction of the scalar potential and, while doing so, we also introduce a Maxwell's field to make the system even more realistic. We will come back to these issues in more detail toward the end of the next subsection.

\subsection{Einstein-Maxwell-scalar system}

With the review of an Einstein-scalar system, we now turn to a more realistic system of an Einstein-Maxwell-scalar system with a Higgs potential.                      
The action is given by \cite{Park:2019amz}
\bea
&&\hspace{-.2in}S=\fr1{\k^2}\int  \sqrt{-g}\;\Big[R-2\L\Big] +\int d^4 x \sqrt{-g}\;  \Big[c_1  R^2+c_2 R_{\m\n}R^{\m\n} +\cdots\Big]  \nn\\
&&\hspace{-.3in}-\fr14 \int \sqrt{-g}\;F_{\m\n}F^{\m\n}  -\int d^4 x \sqrt{-g}\;\Big[|\pa_\mu \psi-iqA_\m \psi|^2 
+{\l}\Big(|\psi|^2+\fr{1}{2\l} \n^2\Big)^2   \Big].
  \la{emsactcasetwo}
 \nn\\
\eea
The metric and scalar field equations are
\bea
&& R_{\m\n}-\L g_{\m\n}
-\fr{\k^2}2g_{\m\n}\Big[ {\l}\Big(|\psi|^2+\fr{1}{2\l} \n^2\Big)^2 -\fr14 F_{\a\b}F^{\a\b}  \nn\\
                       &&\hspace{1in}   +c_1  R^2+(2c_1+c_2)  \nabla^2 R  +c_2  R_{\a\b}R^{\a\b}  +\cdots \Big]  \nn\\
&&+ \k^2\Big[  { -\fr12  \left((\pa_\mu \psi-iqA_\m \psi)(\pa_\n \psi^*+iqA_\n \psi^*)+(\m \leftrightarrow \n)\right)} -\fr12 F_{\m\r}F_\n{}^\r  \nn\\
&&+2c_1  RR_{\m\n}  -(2c_1+c_2)  \nabla_\m \nabla_\n R
                             -2c_2  R_{\k_1\m\n\k_2} R^{\k_1\k_2}+c_2  \nabla^2 R_{\m\n}  +\cdots\Big]  \nn\\
                             &&=0 ,  \la{quanfe}
\eea
\[
{ \nabla^\m F_{\m\n}+iq \psi (\pa_\n+iq A_\n)\psi^*-iq \psi^* (\pa_\n-iq A_\n)\psi  +\cdots=0,}
\]
\bea
\hspace{.4in}(\nabla^\m-iqA^\m)(\nabla_\m-iqA_\m) \psi-\n^2\psi -{2}\l \psi |\psi|^2 +\cdots=0. \qquad \nn
\eea
Below we will obtain, in a series form, a time-dependent solution that settles down to the standard Kerr geometry as the time-dependence fades out. It is thus useful to have the following series expansion of the standard Kerr geometry (note that it is a Kerr geometry but not an (A)dS Kerr for a reason to be explained),
\bea
&& \hspace{.3in}ds^2 = -\Big(1-\fr{2Mz}{1+a^2z^2\cos^2\th} \Big)(dt+a\sin^2\th d\f)^2
\nn\\
&&\hspace{-.3in} +2(dt+a\sin^2\th d\f)\Big(-\fr{dz}{z^2}+a\sin^2\th d\f \Big)+\Big(\fr1{z^2}+a^2\cos^2\th\Big)(d\th^2+\sin^2\th d\f^2),\nn\\
\eea
where the factor in front of $(dt+a\sin^2\th d\f)^2$ can be expanded:
\bea
 1-\fr{2Mz}{1+a^2z^2\cos^2\th} =1-{2 M}{z}+{2 a^2 M \cos ^2\theta }{z^3} -{2 a^4 M \cos ^4\theta }{z^5}+{2 a^6 M \cos ^6\theta }{z^7}+\cdots. \nn\\
\eea
For a time-dependent solution, let us try the following ansatz:
\bea
ds^2&=&-\fr{F(t,z,\th)}{z^2}(dt+a \sin^2\th d\f)^2+2(dt+a \sin^2\th d\f)\Big(-\fr{dz}{z^2}+a\sin^2\th d\f\Big)  \nn\\
  && +\Phi^2(t,z,\th) (d\th^2+\sin^2d\f^2)  \nn\\
&=&-\fr{F(t,z,\theta)}{z^2}dt^2-\fr2{z^2}dtdz+2a\Big(-  \fr{F(t,z,\theta)}{z^2}+1\Big)\sin^2\th\, dtd\f -\fr{2a}{z^2}\sin^2\th dzd\f \nn\\
&&+\Phi^2(t,z,\theta) d\th^2+\Big( -\fr{a^2F(t,z,\theta)}{z^2}\sin^2\th+2a^2\sin^2\th+\Phi^2(t,z,\theta) \Big)\sin^2\th d\f^2. 
\la{Kerrans}
\nn\\
\eea
For the field variables, let us take the following ansatze: for the scalar field,
\bea
\psi(t,z,\th,\phi)&=&\psi_0(t,\th) +\psi_1(t,\th) z+ \psi_2(t,\th)z^2+\psi_3(t,\th)z^3 + ...\nonumber\\
&+&\k^2  \Big[ \psi_0^h(t,\th) +\psi_1^h(t,\th) z+ \psi_2^h(t,\th)z^2+\psi_3^h(t,\th)z^3 + ...\Big].  \nn\\ \la{zetaser}
\eea
For the vector field,\footnote{This form of the ansatz does not cover the charged black hole case. We leave the charged case for the future investigation.}
\bea
A_\m(t,z,\theta,\f)=(0, A_1(t,z,\theta),A_2(t,z,\theta),A_3(t,z,\theta))
\eea
with
\bea
\hspace{-.5in} A_1(t,z,\theta)&=& A_{z0}(t,\th) +A_{z1}(t,\th) z+ A_{z2}(t,\th)z^2+A_{z3}(t,\th)z^3 + ... \nn\\
&+&\k^2  \Big[ A_{z0}^h(t,\th) +A_{z1}^h(t,\th) z+ A_{z2}^h(t,\th)z^2+A_{z3}^h(t,\th)z^3 + ...\Big],\nn\\
A_2(t,z,\theta)&=& A_{\th0}(t,\th) +A_{\th1}(t,\th) z+ A_{\th2}(t,\th)z^2+A_{\th3}(t,\th)z^3 + ... \nn\\
&+&\k^2  \Big[ A_{\th0}^h(t,\th) +A_{\th1}^h(t,\th) z+ A_{\th2}^h(t,\th)z^2+A_{\th3}^h(t,\th)z^3 + ...\Big],\nn\\
A_3(t,z,\theta)&=& A_{\f0}(t,\th) +A_{\f1}(t,\th) z+ A_{\f2}(t,\th)z^2+A_{\f3}(t,\th)z^3 + ... \nn\\
&+&\k^2  \Big[ A_{\f0}^h(t,\th) +A_{\f1}^h(t,\th) z+ A_{\f2}^h(t,\th)z^2+A_{\f3}^h(t,\th)z^3 + ...\Big].\nn\\
\la{vecser}
\eea
For the metric,
\bea
\hspace{-.5in}F(t,z,\th)&=& F_0(t,\th) +F_1(t,\th) z+ F_2(t,\th)z^2+F_3(t,\th)z^3 + ...\nonumber\\
&+&\k^2  \Big[F_0^h(t,\th) +F_1^h(t,\th) z+ F_2^h(t,\th)z^2+F_3^h(t,\th)z^3 + ...\Big], \nn\\
\Phi(t,z,\th)&=&\frac{1}{z}+\Phi_0(t,\th) +\Phi_1(t,\th) z+ \Phi_2(t,\th)z^2+\Phi_3(t,\th)z^3 + ...\nonumber\\
&+& \k^2  \Big[\fr{\Phi_{-1}^h(t,\th)}{z}+\Phi_0^h(t,\th) +\Phi_1^h(t,\th) z+ \Phi_2^h(t,\th)z^2+\Phi_3^h(t,\th)z^3 + ...\Big] \nn\\
   \la{1stans}
\eea
where the modes with superscript `$h$' represent the quantum modes. The quantum corrections of the metric imply a deformation of the geometry by quantum effects \cite{Park:2013rm}. The cosmological constant $\L$ is set to $\L= \L_0+\hbar \k^2 \L_1 $ with vanishing $\L_0$, $\L_0=0$, to prevent occurrence of the undesirable feature noted for the Einstein-scalar system. 
One can consider the first several $z$-powers. For each $z$-power, expand the coefficients to the first order of $\hbar$.
Our analysis yields the following results: for the classical modes,
\bea
&&\psi_0(t,\th)=\psi_0,\quad \psi_0 \psi_0^*=-\fr{\n^2}{2\l},\quad\nn\\
&&  \psi_1(t,\th)=0,\quad \psi_2(t,\th)=0,\quad \psi_3(t,\th)=0,\quad  \nn\\
&&A_{z0}(t,\th)=0,\quad A_{\th0}(t,\th)=0,\quad A_{\f0}(t,\th)=0, \nn\\
&& A_{z1}(t,\th)=0,\quad A_{\th1}(t,\th)=0,\quad A_{\f1}(t,\th)=0, \nn\\
&& A_{z2}(t,\theta )=0,\quad A_{\th2}(t,\th)=0,\quad A_{\f2}(t,\th)=0,\nn\\
&&  A_{\th3}(t,\th)=0,\quad  A_{\f3}(t,\th)=0,\quad  A_{\f4}(t,\th)=0,\quad \nn\\
&&  F_0(t,\th)=0,\quad  F_1(t,\theta)=0,\quad F_2(t,\theta)=1 ,\quad\nn\\
&& F_3(t,\theta) =\mathrm{const},\quad F_4(t,\theta) =0,\nn\\
&& \Phi _{-1}(t,\theta )=1,\quad \Phi _0(t,\theta )=0,\quad  \Phi _1(t,\theta )=\fr12 a^2 \cos^2\th,\quad 
\nn\\
&&\Phi _2(t,\theta )=0,\quad \Phi_3(t,\theta)=-\frac{1}{8} a^4 \cos ^4\theta; \nn\\
\la{moderes}
\eea
for the quantum modes,
\bea
&&\pa_t \psi_0^h(t,\th)=0,\quad  \psi_0^{h*}(t,\th)= \frac{\nu ^2 \psi^h _0(t,\theta )}{2 \lambda  \psi _0^2},\quad \psi^{h*}_1(t,\theta )=\frac{\nu ^2 \psi^h_1(t,\theta )}{2 \lambda  \psi _0^2} , \nn\\
&&  \pa_t\psi^{h}_1(t,\theta ) =0,\quad \pa_t\psi^h_2(t,\theta )=0,\quad   \psi^{h*}_2(t,\theta )=\frac{\nu ^2 \psi^h_2(t,\theta )}{2 \lambda  \psi _0^2},\quad\nn\\
&&\psi^{h*}_3(t,\theta)=\frac{\nu ^2 \psi^h_3(t,\theta )}{2 \lambda  \psi _0^2},  \nn\\
&&A^h_{z0}(t,\theta )=-\frac{i \psi^h_1(t,\theta )}{q \psi _0},\quad A_{\th0}^h(t,\th)=-\frac{i\pa_\th \psi^h_0{}(t,\theta )}{q \psi _0},\quad A_{\f0}^h(t,\th)=0,\quad
 \nn\\
&& A_{\th1}^h(t,\th)=-\frac{i \pa_\th\psi^h_1(t,\theta )}{q \psi _0},\quad A_{\f1}^h(t,\th)=0,\quad A_{z1}^h(t,\theta)=-\frac{2 i \psi^h_2(t,\theta )}{q \psi _0},\quad
\nn\\
&& A_{z2}^h(t,\theta)=-\frac{3 i \left(2 \lambda  \psi _0^2 \psi^{h*}_3(t,\theta )+\nu ^2 \psi^h_3(t,\theta )\right)}{2 \nu ^2 q \psi _0} ,\quad A^h_{\f2}(t,\theta )=0,
\nn\\
&&A_{\th2}^h(t,\th)=-\frac{i \left(\nu ^2 q^2 \pa_\th\psi^h_2(t,\theta )+\lambda \pa_t\pa_\th \psi^h_1(t,\theta )\right)}{\nu ^2 q^3 \psi _0},\quad A_{\f3}^h(t,\th)=0 ,\nn\\
&&F_0^h(t,\th)=-\frac{1}{3}  \Lambda_1,\quad  F_1^h(t,\th)=-2\pa_t \Phi_{-1}^h,\quad    \nn\\
&&F^h_2(t,\theta)=-\frac{5}{3}a^2  \Lambda_1 \cos ^2\theta +2 \Phi^h_{-1}(t,\theta )-2 (\cos 2 \theta +2) \csc\theta \sec\theta \, \pa_\th\Phi^h_{-1}(t,\theta ) ,\nn\\
&&\nn\\
&&\pa_\th^2\Phi^h_{-1}(t,\theta )=\frac{1}{4} \Big(-\cot ^2\theta  \left[3 F_3(t,\theta ) \pa_t\Phi^h_{-1}(t,\theta )+a^2  \Lambda_1 (\cos 2 \theta +3)\right]\nn\\
&&\hspace{1.5in}-2 (\cos 2 \theta +3) \csc \theta  \sec \theta  \,\pa_\th\Phi^h_{-1}(t,\theta )\Big) ,\nn\\
&& \pa_t\Phi_0^h(t,\theta )=-2\Phi_{-1}^h(t,\theta )+2\cot\th \,\pa_\th\Phi_{-1}^h(t,\theta),\quad \nn\\
&&\pa_\th\Phi^h_0(t,\theta )=-2 a^2 \sin \theta  \cos \theta  \,\pa_t\Phi^h_{-1}(t,\theta ) ,\nn\\
&& \Phi_1^h(t,\theta )=-\fr32 a^2 \cos^2\th \,\Phi_{-1}^h(t,\theta ),\quad \Phi^h_2(t,\theta)=-\frac{1}{2} a^2 \cos^2\theta \, \Phi^h_0(t,\theta),\quad \nn\\
&&\Phi_3^h(t,\theta)=\frac{7}{8} a^4 \cos^4\theta \,\Phi_{-1}^h(t,\theta).
\eea
Several remarks are in order. 
Although the field equations are more complex and entangled, there exists, as in \cite{Nurmagambetov:2018het}, a robust pattern in the manner in which the mode relationships above are obtained. The lowest $\hbar$- and $\k$- order terms in the action \rf{emsactcasetwo} are important in determining the building blocks of the higher modes. Some of the leading $\hbar$-correction parts introduce additional constraints among the classical modes - which is thus of ``order-1" effect, and thereby qualitatively change the classical part of the solution. 
As the higher-order terms (that are not explicitly shown in \rf{emsactcasetwo}) are added to the action \rf{emsactcasetwo}  and thus to the field equations, their effects are limited to the newly introduced higher modes that are then determined in terms of the lower modes. Some mode results in \rf{moderes} deserve specific comments: one particularly novel feature is that, just as in the Einstein-scalar case analyzed in \cite{Nurmagambetov:2018het}, the classical modes of the matter fields (i.e., the scalar field and Maxwell's field) are removed by the quantum-level constraints. There are differences as well. One of the differences is that, unlike in \cite{Nurmagambetov:2018het}, in which the mode $\Phi _0(t,\th)$ is not constrained, it is constrained to vanish  here. The field equations constrain $F_3$ to be a constant; with the requirement that the solution settles down to the usual Kerr geometry, it is determined to be $F_3=-2M$.

The results above have been obtained by setting $\L_0$ to $\L_0=0$ from the beginning. Let us clarify this. We suspect that the undesirable feature noted in the Einstein-scalar case should be due to the inadequacy of the ansatz \rf{Kerrans} for an (A)dS case. 
In the literature, the (A)dS Kerr solution is long known in the Boyer-Lindquist coordinates. The first step in proper handling of a time-dependent (A)dS Kerr solution should be to write down the ansatz based on the (A)dS Kerr solution in the Eddington-Finkelstein-type coordinates that we have employed. (More on this in the conclusion.)

\section{Near-horizon dynamics}

The upshot of the previous section is that the essential quantum-level  physics can be captured by the action eq. \rf{emsactcasetwo}, 
and the system admits the following time-dependent solution
\bea
\hspace{-.5in}g_{\m\n}=\left(
\begin{array}{cccc}
 -\frac{F(t,z,\theta )}{z^2} & -\frac{1}{z^2} & 0 & a \left(1-\frac{F(t,z,\theta )}{z^2}\right) \sin ^2\theta  \\
 -\frac{1}{z^2} & 0 & 0 & -\frac{a \sin ^2\theta }{z^2} \\
 0 & 0 & \Phi^2(t,z,\theta ) & 0 \\
 a \left(1-\frac{F(t,z,\theta )}{z^2}\right) \sin ^2\theta  & -\frac{a \sin ^2\theta }{z^2} & 0 & 2 a^2 \sin ^4\theta -\frac{a^2 F(t,z,\theta ) \sin ^4\theta }{z^2}+\Phi^2(t,z,\theta ) \sin ^2\theta  \\
\end{array}
\right), \nn\\
\eea
\[
A_\m(t,z,\theta,\f)=(0, A_1(t,z,\theta),A_2(t,z,\theta),A_3(t,z,\theta)),
\]
\[
\psi(t,z,\th,\phi)=\psi(t,z,\th),
\]
with each field component expanded in terms of the modes
that satisfy the relationships given in \rf{moderes}. In this section, we complete the rest of the steps of the energy computation.

To compute the energy in the leading order in $\hbar$ and $\k$, it suffices to compute only the classical geodesic, a feature shared by the Einstein-scalar system considered in the previous work, \cite{Nurmagambetov:2018het}. Since the time-dependence of the classical part of the solution is removed by the quantum effects, the classical geometry is that of the Kerr. 
It turns out that it is the boundary modes that are the building blocks of the time-dependence and represent the deformations. They also take a part in the trans-Planckian energy.

In section 4.1, we review the computation of the geodesic in the Kerr background.  In section 4.2, we consider reexpansion of the solution around the classical location of the event horizon. In the analysis analogous to that in \cite{Nurmagambetov:2018het}, we show that the energy measured by an infalling obsever is trans-Planckian. 
 What we called the ``horizon quantum modes" in \cite{Nurmagambetov:2018het} leads to the trans-Planckian energy. In section 4.3, we comment on the boundary conditions.

\subsection{Four-velocity of an infalling observer}

One of the ingredients needed to compute the local energy measured by an infalling observer is the four-velocity vector (see e.g. \cite{Lowe:2013zxa} \cite{Park:2017dib} \cite{Nurmagambetov:2018het}). As seen in the previous section, the time-dependent pieces of the classical part of the quantum-level solution become constrained to vanish: the time-dependent part of the solution for the field equations is only the quantum correction piece. This implies that the classical part of the stress-energy is that of a Kerr geometry. Since the stress-energy tensor vanishes for a Kerr geometry, one can use the geodesic analysis of Kerr spacetime in order to compute the leading quantum-gravitational correction of the energy.

Let us review the geodesic analysis of Kerr spacetime \cite{Carter:1968rr}.
The metric admits two Killing vectors:
\be
k^\m_t=(1,0,0,0),\qquad k^\m_\vf=(0,0,0,1),
\la{KillingKN}
\ee
which leads to two integrals to the geodesic equations: the energy
\be
E=-g_{\m\n}k^\m_t U^\n,
\la{EKN}
\ee
and angular momentum projection
\be
l=g_{\m\n} k^\m_\vf U^\n ,
\la{lphiKN}
\ee
where $U^\n$ is the four-velocity $U^\m\equiv \fr{dx^\m}{d\l}$ with $\l$ being the proper-time parameter along the geodesic.
It is normalized according to 
\be
g_{\m\n} \fr{dx^\m}{d \l} \fr{dx^\n}{d \l}=-\m^2 .
\la{massshell}
\ee
{The time-like and light-like geodesics correspond to $\m=1,0$, respectively.}
One can show that the four-velocity components are given by (the dot represents $\fr{d}{d\l}$)
\be
 \dot{\vf} 
=\fr1{\fr1{z^2}+a^2\cos^2\th}\Big[\left(aE+\fr{l}{\sin^2\th}\right)-a\triangle^{-1}(P+\sqrt{R})\Big],
\la{SigmaDotphiz}
\ee
\be
\dot{t}=\fr1{\fr1{z^2}+a^2\cos^2\th}\Big[-a(l+a E \sin^2\th)+\left(\fr1{z^2}+a^2\right)\triangle^{-1}(P+\sqrt{R} ) \Big],
\la{SigmaDotuz}
\ee
\[
\dot{z}=-\fr{z^2}{\fr1{z^2}+a^2\cos^2\th}\sqrt{R} \,,
\]
\be
 \dot{\th}=\fr1{\fr1{z^2}+a^2\cos^2\th}\sqrt{\Th}
\la{SigmaDotthz}
\ee
with
\be
\triangle=a^2+z^{-2}-\fr{2M}{z},
\la{triangl}
\ee
\be
P=al+(a^2+z^{-2})E ,
\la{Pdef}
\ee
\be
\Th=\mathcal{K}-(l+Ea)^2-\cos^2 \th \left[a^2(\m^2-E^2)+\fr{l^2_0}{\sin^2\th} \right],
\la{Thdef}
\ee
\be
R=P^2-\triangle \left(\mathcal{K}+\fr{\m^2}{z^2} \right),
\la{Rdef}
\ee
where $\mathcal{K}$ is another integral of motion called the Carter constant.

\subsection{Trans-Planckian energy near horizon}

The energy density as measured by a free-falling observer is given by
\bea
\r\equiv T_{\m\n}U^\m U^\n.  \la{2dse}
\eea
The stress-energy tensor\footnote{See \cite{Candelas:1980zt,Birrell,Frolov,Mukhanov} for reviews on the quantum-level stress tensor.} is obtained by taking the functional derivative of the matter part of the action with respect to the metric: 
\bea
{ T_{\m\n}} &=&- \fr2{\k^2}\L g_{\m\n}+g_{\m\n}\Big[-|\pa_\r \psi-iqA_\r \psi|^2 -{\l}\Big(|\psi|^2+\fr{1}{2\l} \n^2\Big)^2 -\fr14 F_{\r\s}^2  \nn\\
 &&\hspace{.6in} +\hbar \Big(c_1R^2-(4c_1+c_2)\nabla^2 R  +c_2 R_{\r\s}R^{\r\s}\Big)+\cdots \Big] 
\la{quanset}
\eea
\[\hspace{-.3in} +\Big[ { \left((\pa_\mu \psi-iqA_\m \psi)(\pa_\n \psi^*+iqA_\n \psi^*)+(\m \leftrightarrow \n)\right)}+  F_{\m\r}F_\n{}^\r
\]
\[
\hspace{.2in}-2\hbar \Big(2c_1 RR_{\m\n}  -(2c_1+c_2) \nabla_\m \nabla_\n R
-2c_2 R_{\k_1\m\n\k_2} R^{\k_1\k_2}+c_2\nabla^2 R_{\m\n}\Big)  +\cdots \Big].  
\]
For the leading-order energy correction\footnote{The leading correction is of second power in $\hbar$ and of inverse second power in $\k$ (after the $\k$-rescaling of the matter fields to be discussed below).}, one can use the classical form of the stress-energy tensor,
\bea
&& \hspace{-.2in}{ T_{\m\n}} = - \fr2{\k^2}\L g_{\m\n}+g_{\m\n}\Big[-|\pa_\r \psi-iqA_\r \psi|^2 -{\l}\Big(|\psi|^2+\fr{1}{2\l} \n^2\Big)^2 -\fr14 F_{\r\s}^2   \Big] \nn\\
&&\hspace{.2in} +  \left[(\pa_\mu \psi-iqA_\m \psi)(\pa_\n \psi^*+iqA_\n \psi^*)+(\m \leftrightarrow \n)\right]+  F_{\m\r}F_\n{}^\r.
\eea 
Note that although the classical form of the stress-energy tensor is used, the full quantum-level solution is to be substituted into the tensor. An additional simplifying feature is that the terms with $g_{\m\n}$ are bound by the geodesic normalization, $U_\m U^\m=-\m^2$, and thus unimportant.\footnote{The four-velocity reviewed in section 4.1 was at the classical level and, in particular, so is the normalization condition. However, in anticipation of the full quantum-level normalization condition, we omit the terms with $g_{\m\n}$ from the present leading-order energy correction computation.} 
Thus the leading-order quantum correction to the classical energy is given by
\bea
\r &\sim& \Big( [(\pa_\mu \psi-iqA_\m \psi)(\pa_\n \psi^*+iqA_\n \psi^*)+(\m \leftrightarrow \n)]+  F_{\m\r}F_\n{}^\r\Big)
U^\m U^\n.  \la{rhorel} \nn\\ 
\eea
Pole terms - which later yield trans-Planckian scaling - arise from the scalar or vector kinetic term above. More specifically, the $\pa_t{\y}\pa_t{\y}^*\dot{t} \dot{t}$ and $F_{t\r}F_t{}^\r \dot{t} \dot{t}$ terms produce the pole terms\footnote{As a matter of fact, $\dot\vf$ has a pole too. We will focus on $\dot{t}$.}: $\dot{t}$ scales as $\dot{t} \sim \fr{1}{z-z_{EH}}$ where the classical horizon $z_{EH}$ is located at the vanishing of $a^2+z^{-2}-{2M}z^{-1}$. 

Since the four-velocity has a pole at the classical location of the event horizon $z=z_{EH}$, a more transparent understanding of the behavior of the matter fields near the horizon can be gained by considering reexpansion of the $z$-series solution in 
\be
Y\equiv z-z_{EH}\,. 
\ee
Let us suppose the following form of expansion of the matter fields around $z_{EH}$\footnote{At least to the orders analyzed, the time-dependence of the classical parts of the matter fields is absent. In \cite{Nurmagambetov:2018het}, the similar feature was checked to remain true to all orders in $z$, not just to the first several orders. That was done by solving the field equations with the expansion given in eq. \rf{vecYser}. Although the corresponding task for the present system turns out to be too involved, we expect that the feature remains true.}:  
\bea
\psi(t,z,\th)&=& \tilde{\psi}_0(t,\th) +  \tilde{\psi}_1(t,\th) Y+ \tilde{\psi}_2(t,\th)Y^2+\tilde{\psi}_3 Y^3 +\cdots\nn\\
&&+\k^2\Big[\tilde{\psi}_0^h(t,\th) +  \tilde{\psi}_1^h(t,\th) Y+ \tilde{\psi}_2^h(t,\th)Y^2+\tilde{\psi}_3^h Y^3 +\cdots \Big] , \nn\\
\hspace{-.5in} A_1(t,z,\theta)&=& \At_{z0}(t,\th) +\At_{z1}(t,\th) Y+ \At_{z2}(t,\th)Y^2+\At_{z3}(t,\th)Y^3 + ... \nn\\
&&+\k^2  \Big[ \At_{z0}^h(t,\th) +\At_{z1}^h(t,\th) Y+ \At_{z2}^h(t,\th)Y^2+\At_{z3}^h(t,\th)Y^3 + ...\Big] ,\nn\\
A_2(t,z,\theta)&=& \At_{\th0}(t,\th) +\At_{\th1}(t,\th) Y+ \At_{\th2}(t,\th)Y^2+\At_{\th3}(t,\th)Y^3 + ... \nn\\
&&+\k^2  \Big[ \At_{\th0}^h(t,\th) +\At_{\th1}^h(t,\th) Y+ \At_{\th2}^h(t,\th)Y^2+\At_{\th3}^h(t,\th)Y^3 + ...\Big] ,\nn\\
A_3(t,z,\theta)&=& \At_{\f0}(t,\th) +\At_{\f1}(t,\th) Y+ \At_{\f2}(t,\th)Y^2+\At_{\f3}(t,\th)Y^3 + ... \nn\\
&&+\k^2  \Big[ \At_{\f0}^h(t,\th) +\At_{\f1}^h(t,\th) Y+ \At_{\f2}^h(t,\th)Y^2+\At_{\f3}^h(t,\th)Y^3 + ...\Big].\nn\\
\la{vecYser}
\eea
The `tilded' modes will be given as sums of the original modes. To the orders that we have checked in section 3, all of the classical modes (except $\psi_0$, which is irrelevant for the energy computation) vanish; because of this the mode expansion above gets simplified to
\bea
\psi(t,z,\th)&=& \tilde{\psi} +\k^2\Big[\tilde{\psi}_0^h(t,\th) +  \tilde{\psi}_1^h(t,\th) Y+ \tilde{\psi}_2^h(t,\th)Y^2+\tilde{\psi}_3^h Y^3 +\cdots \Big] ,  \nn\\
\hspace{-.5in} A_1(t,z,\theta)&=& 
\k^2  \Big[ \At_{z0}^h(t,\th) +\At_{z1}^h(t,\th) Y+ \At_{z2}^h(t,\th)Y^2+\At_{z3}^h(t,\th)Y^3 + ...\Big] ,\nn\\
A_2(t,z,\theta)&=& \k^2  \Big[ \At_{\th0}^h(t,\th) +\At_{\th1}^h(t,\th) Y+ \At_{\th2}^h(t,\th)Y^2+\At_{\th3}^h(t,\th)Y^3 + ...\Big] ,\nn\\
A_3(t,z,\theta)&=& \k^2  \Big[ \At_{\f0}^h(t,\th) +\At_{\f1}^h(t,\th) Y+ \At_{\f2}^h(t,\th)Y^2+\At_{\f3}^h(t,\th)Y^3 + ...\Big].\nn\\
\la{vecYsersim}
\eea

Before getting to the final-stage energy analysis, let us note that rescaling of the matter fields is necessary for correct $\k$-scaling of various physical quantities, including the energy. The fact that the matter part of the action comes at higher order of $\k^2$ implies \cite{Chadburn:2013mta}\cite{Nurmagambetov:2018het} that the solution generically takes the form of  
\bea
\psi=\fr{\xi}{\k}\quad,\quad A_m=\fr{a_m}{\k}, \quad m=1,2,3\,,
\eea
where $\xi, a_m$ represents the rescaled scalar and vector fields; they will have series expansions - which are similar to those in eq. \rf{vecYser} - in terms of the modes with tildes. In particular, the modes $(\tilde{\xi}^h_0(t,\th),\at_{z0}^h(t,\th), \at_{\th0}^h(t,\th), \at_{\f0}^h(t,\th))$ play an important role in the energy, as we will now see. 
The location of the horizon at the quantum level, $z_{EH}^h$, (whose precise determination we do not pursue in the work) will take the form of
\bea
z_{EH}^h=z_{EH}+ {\cal O}(\k^{2})
\eea 
and this implies
\bea
\dot{t}\sim {\cal O}(\k^{-2})
\eea
at $z=z_{EH}^h$. With this scaling one gets, for the leading behavior of $\r$,
\bea
T_{\m\n} \;U^\m U^\n \sim \fr{\k^2 f(\tilde{\xi}^h_0,\at_{z0}^h, \at_{\th0}^h, \at_{\f0}^h)}{\k^4} \sim \fr1{\k^2} \,,
\eea
where $f(\tilde{\xi}^h_0,\at_{z0}^h, \at_{\th0}^h, \at_{\f0}^h)$ is a quantity that is proportional to $T_{00}$. A direct calculation yields\footnote{Similarly, one gets
\bea
T_{33} &=&  2 q^2 \tilde{\xi}_0 \tilde{\xi}_0^* (\tilde{a}^h_{\f0})^2 
     +\frac{z_{EH}^2}{a^2 z_{EH}^2 \cos^2\theta +1}
\Big(  a^2 \sin^2\theta (\pa_t \tilde{a}^h_{\f0})^2+ (\pa_\th \tilde{a}^h_{\f0})^2  \nn\\
&& \hspace{1in} -2 \pa_t \tilde{a}^h_{\f0} \tilde{a}^h_{\f1} [a^2 z_{EH}^2 \cos^2\theta +a^2 z_{EH}^2 \sin^2\theta +1]  \nn\\
&&\hspace{.8in} +z_{EH}^2 (\tilde{a}^h_{\f1})^2 [a^2 z_{EH}^2 \cos^2\theta +a^2 z_{EH}^2 \sin^2\theta -2 M z_{EH}+1]
 \Big).
\eea
}
\bea
T_{00} &=& \frac1{\sin^2\theta  \left(a^2 z_{EH}^2 \cos^2\theta +1\right)} 
\Big({a^2 z_{EH}^6 \sin^4\theta  (\pa_t \tilde{a}^h_{z0})^2+z_{EH}^2 (\pa_t \tilde{a}^h_{\f0})^2} \nn\\
&&\hspace{-.7in} +\sin^2\theta  \Big[z_{EH}^4 (\pa_t \tilde{a}^h_{z0})^2 \left(a^2 z_{EH}^2 \cos^2\theta -2 M {z_{EH}}+1\right)   +2 \pa_t \tilde{\xi}^{h*}_0 \pa_t \tilde{\xi}^h_0 \left(a^2 z_{EH}^2 \cos^2\theta +1\right) \nn\\
	&&\hspace{1in}+2 a z_{EH}^4 \pa_t \tilde{a}^h_{z0} \pa_t \tilde{a}^h_{\f0}+z_{EH}^2 (\pa_t \tilde{a}^h_{\th0})^2\Big] \Big) +\cdots  ,
\eea
where the explicitly shown terms represent the expression inside the parentheses in \rf{rhorel}.

\subsection{On the boundary conditions}

Dirichlet boundary conditions have been widely considered in quantum and gravitational field theories. The recent works show, however, that a complete description of a gravitational system requires extension of the Hilbert space by including other boundary conditions \cite{Park:2019amz}. Let us examine the mode expansions \rf{zetaser}, \rf{vecser}, and \rf{1stans}. The presence of the dynamic boundary modes such as
$\Phi_{-1}^h(t,\th),\Phi_0^h(t,\th))$ implies that the solution satisfies a certain Neumann-type boundary condition but not a Dirichlet boundary condition. One noteworthy point is that the boundary condition at the asymptotic infinity closely controls what's happening at the event horizon, as one can see by examining the reexpansions eq. \rf{vecYsersim} (and those of the metric fields).

In certain circumstances such as a black hole merger, a fixed boundary condition is considered at the horizon. The most widely used one is a perfect-infall boundary condition in the context of the quasi-normal modes. However, the quantum effects obtained in \cite{Nurmagambetov:2018het} and the present work seem to suggest more general and inclusive boundary conditions as the natural ones at the horizon.  
The presence of the aforementioned quantum boundary modes will imply that both transmitted and reflected waves will be present.  
When the time-dependence fades out, the boundary modes become die out and a Dirichlet boundary condition naturally arises.

\section{Conclusion}

In this work we have considered a time-dependent quantum-corrected black hole solution of an Einstein-Maxwell-scalar with a Higgs potential. We have computed the near-horizon energy measured by an infalling observer. The analysis consists of three components: computation of the quantum action (for which we referred to the previous works) and its time-dependent solution, computation of the four-velocity of the observer, as well as evaluation of the energy after expansion around the location of the horizon. As in the previous works \cite{Park:2017dib}\cite{Nurmagambetov:2018het}, a trans-Planckian energy has resulted. Concerning the physical origin of the trans-Planckian energy behavior, the loop effects become important near the horizon. 

The power of the previous works and that of the present is the generality of the analyses and the quantitative conclusions drawn from them. In other words, there exists a robust pattern in determination of the higher order modes in terms of the lower ones: the solutions are built out of several lower modes for which only the first two leading quantum correction terms in the action, $R^2, R_{\r\s}R^{\r\s}$, are important.

The present work is motivated in part by recent developments in astrophysics, in particular, ultra high energy cosmic rays (UHECRs). Although further work is required, the recent observations indicate active galactic nuclei (AGNs) - the central supermassive black holes of active galaxies - as the candidates of the steady sources of the UHECRs. The relatively new paradigm in the field is the view of the black hole at the center of an AGN as a highly efficient engine that converts the gravitational infall energy into outgoing radiation energy. The quantum gravitational effects may well be the mechanism of generating the energy that feeds various outgoing radiation, especially the UHECRs. 
Being a more realistic system, we expect that the result of the present work will have applications to the physics of $\g$AGNs and UHECRs.

\vspace{.2in}
Before embarking on such an enterprise, there are several more urgent and immediate issues to settle. We present them as near-future directions: 
\vspace{.1in}

An undesirable feature of the solution of the free scalar sector has motivated the introduction of the potential for the scalar sector and consideration of the vanishing classical part of the cosmological constant. Because of the vanishing cosmological constant, our time-dependent solution settles down to the usual Kerr black hole as opposed to a dS/AdS Kerr black hole. In order to relax the restriction  $\L_0=0$, one should first consider the (A)dS-Kerr solution in the present Eddington-Finkelstein-type coordinates. The (A)dS-Kerr solution was long known in the Boyer-Lindquist coordinates. It appears that the solution has recently been converted in \cite{Lake:2015xca} into the Eddington-Finkelstein-type coordinates that we have employed in the present work.

Another relatively urgent direction is a charged black hole case. Taking the temporal gauge, the solution obtained in this work cannot cover the time-dependent extension of a charged black hole, since the standard charged black hole has a nonzero $A_0$ component. It will be of some interest to study how to incorporate the charged case. Presumably a different gauge choice will have to be made.

Still another direction that lies directly ahead of the path of the present work is curved space electrodynamics. The present setup lays necessary foundations for carrying out quantum-gravitational scalar electrodynamics. In particular, it should be possible to compute the quantum-gravitational Poynting vector, which should be useful in comparing with the present and future observations made for astronomical black holes.

\vspace{.3in}

\newpage


\end{document}